\documentclass[]{book}

\usepackage{bm}
\usepackage{url}
\usepackage{hyperref}
\usepackage{graphicx}
\usepackage{caption}
\usepackage{subcaption}
\usepackage{color}
\usepackage[utf8]{inputenc}
\usepackage{makeidx}
\usepackage{cite}
\usepackage{multirow}
\usepackage{todonotes}
\usepackage{colortbl}
\usepackage{amsmath}
\usepackage{amssymb}

\begin{document}

\title{The Utility of Phase Models in Studying Neural Synchronization}
\author{Youngmin Park, Stewart Heitmann, and G. Bard Ermentrout\\ Department of Mathematics\\University of Pittsburgh\\ Pittsburgh PA 15260}

%


\maketitle


\section{Introduction}
Synchronization of neural activity is a ubiquitous phenomenon in the brain that has been associated with many cognitive functions \cite{fries_2005,schnitzler_gross_2005,fell_axmacher_2011} and is inferred from macroscopic electrophysiological recordings (LFP, EEG, MEG). These physiological recordings represent the aggregate rhythmic electrical activity in the cortex \cite{nunez_srinivasan_2006}. To study synchrony, we consider only tonically firing neurons, allowing us to study synchrony solely in terms of spike times.

Weakly coupled oscillator theory \cite{schwemmer_lewis2012}  provides a mechanistic description of synchronization rates and stability. We use this theory to predict and explain synchronization in two types of membranes: Class I membranes, which are characterized by the onset of oscillations that have nonzero amplitude and arbitrarily low frequency, and Class II membranes, which are characterized by the onset of oscillations at non-zero frequency with arbitrarily small amplitude. In terms of dynamics, Class II is associated with a Hopf bifurcation \cite{brown_etal_2004} and Class I is associated with a saddle-node limit cycle (SNLC) bifurcation \cite{ermentrout_1996}.  (Class I,II are also called Type I,II, but to avoid confusion between the classification of PRCs, we will use Class I,II to describe the neuronal dynamics.)  

Reciprocally coupled neurons can synchronize their spiking according to how they respond to incoming spikes. The timing of spike events in a tonically firing neuron can be represented mathematically as the phase of an oscillator. The impact of incoming spikes on that neuron can thus be reduced to perturbations to the phase of an oscillator. How the perturbations advance or delay the phase is quantified by the phase response curve (PRC) and is typically measured directly from the neuron.

\begin{figure}[h!]
 \includegraphics[width=\textwidth]{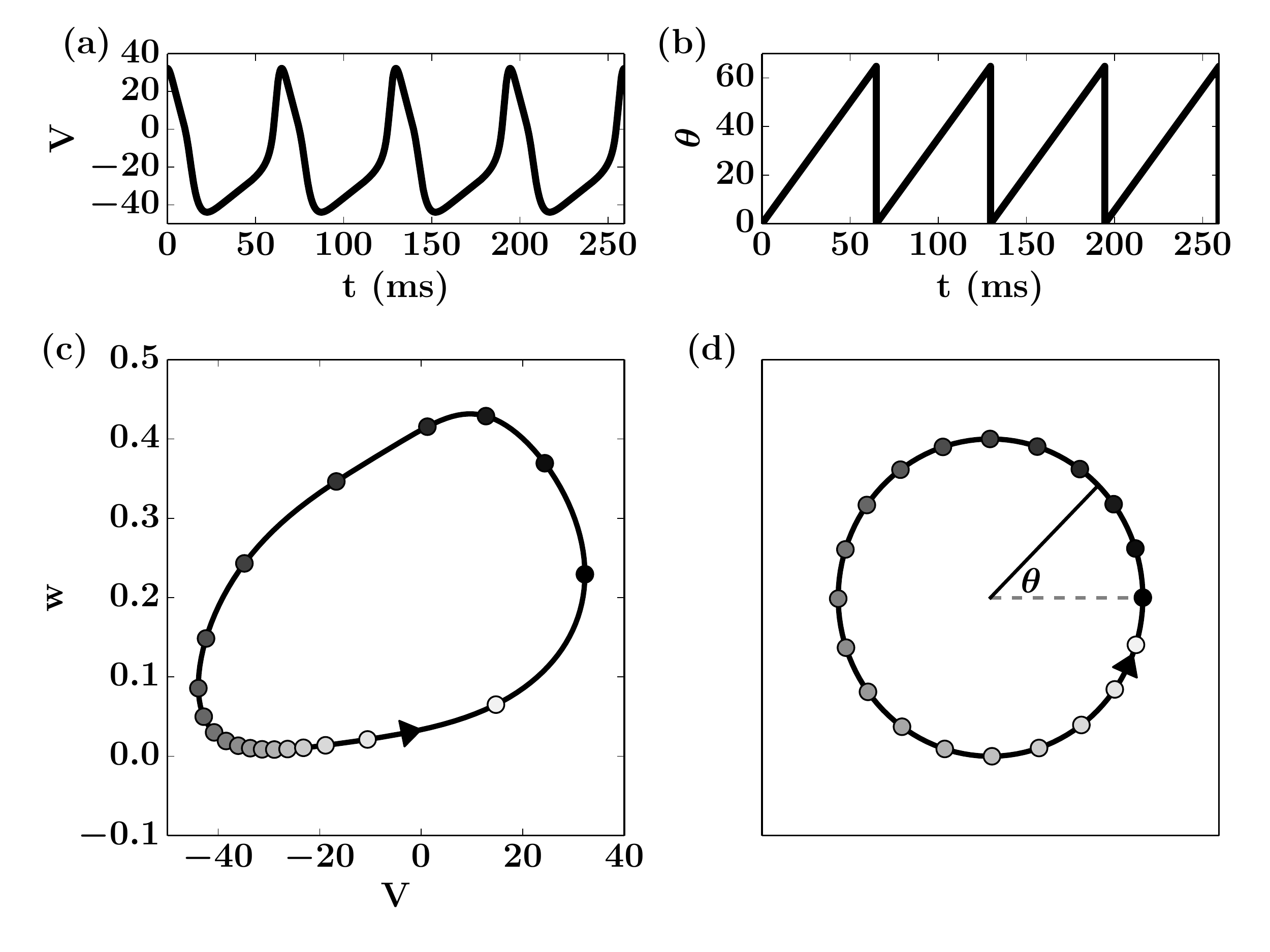}
 \caption{Phase approximation of tonic spiking of the Morris-Lecar model. (a) Membrane potential over time. (b) The phase as a functon of time. (c) Phase plane. The black loop represents the limit cycle with the arrow denoting direction of flow. The open circles represent equally spaced phase values in time. The phase transitions from 0 (black) to $T$ (white). (d) The phase model.} 
 \label{fig:phase_plane}
\end{figure}

In Fig.~\ref{fig:phase_plane}(a), we show repetitive spiking in the Morris-Lecar model, a  simple planar conductance-based model that was originally developed to explain molluscan muscle fibers \cite{morris_lecar_1981,jrbe98}.
The corresponding phase of the spike train is shown in Fig.~\ref{fig:phase_plane}(b). By plotting the voltage and gating variables of the spike train as a parametric curve, we attain Fig.~\ref{fig:phase_plane}(c), the phase space representation of the model. The closed orbit that is shown is both periodic and attracting and therefore a limit cycle, which we denote $\gamma(t)$.

The phase representation in Fig.~\ref{fig:phase_plane}(c,d) is achieved by parameterizing the $T$-periodic limit cycle $\gamma$ by a parameter $\theta \in [0,T)$. This formalism is standard in mathematical neuroscience.

\section{Derivation of the Phase Model}

The phase representation of a neuron allows for a substantial reduction in dimensionality of the system that is particularly useful when studying many coupled neurons in networks. All the complex biophysics, channels, ions, and synaptic interactions are reduced to a set of $N$ coupled phase-models where $N$ is the number of neurons in the network.  The task at hand is to derive how the phases interact when coupled into a network. This simplification to phase comes with some assumptions that we will outline in the ensuing paragraphs.
 
For a generic membrane model, we assume the existence of a $T$-periodic limit cycle, $\gamma(t) = \gamma(t+T)$, satisfying a system of ordinary differential equations,
\begin{equation}\label{eq:odemain}
 \frac{d\vec X}{dt} = \vec F(\vec X),
\end{equation}
where $\vec X \subset \mathbb{R}^n$ and $\vec F: U \subset \mathbb{R}^n \rightarrow \mathbb{R}^n$ is a sufficiently differentiable function. The limit cycle is attracting. In neural models, the limit cycle represents the dynamics of a spiking neural membrane (for example, when injected with a currect sufficient to induce repetitive firing), where one dimension typically represents the membrane voltage and the other dimensions represent recovery variables. 

The phase of the limit cycle $\gamma(t)$ is a function $\theta(t) \in [0,T)$. The phase can be rescaled into any other interval -- common choices include $[0,1)$ and $[0,2\pi)$ -- but we choose $[0,T)$ for convenience. In addition, we choose the phase to satisfy
\[
\frac{d\theta}{dt} = 1.
\]
This choice is a substantial yet powerful simplification of the neural dynamics, which allows us to study deviations from this constant rate, and in turn provide information about spike delays or advances. We account for different models with different spiking frequencies by rescaling time appropriately.

\subsection{Isochrons}
\begin{figure}[h!]
\centering
 \includegraphics[width=.75\textwidth]{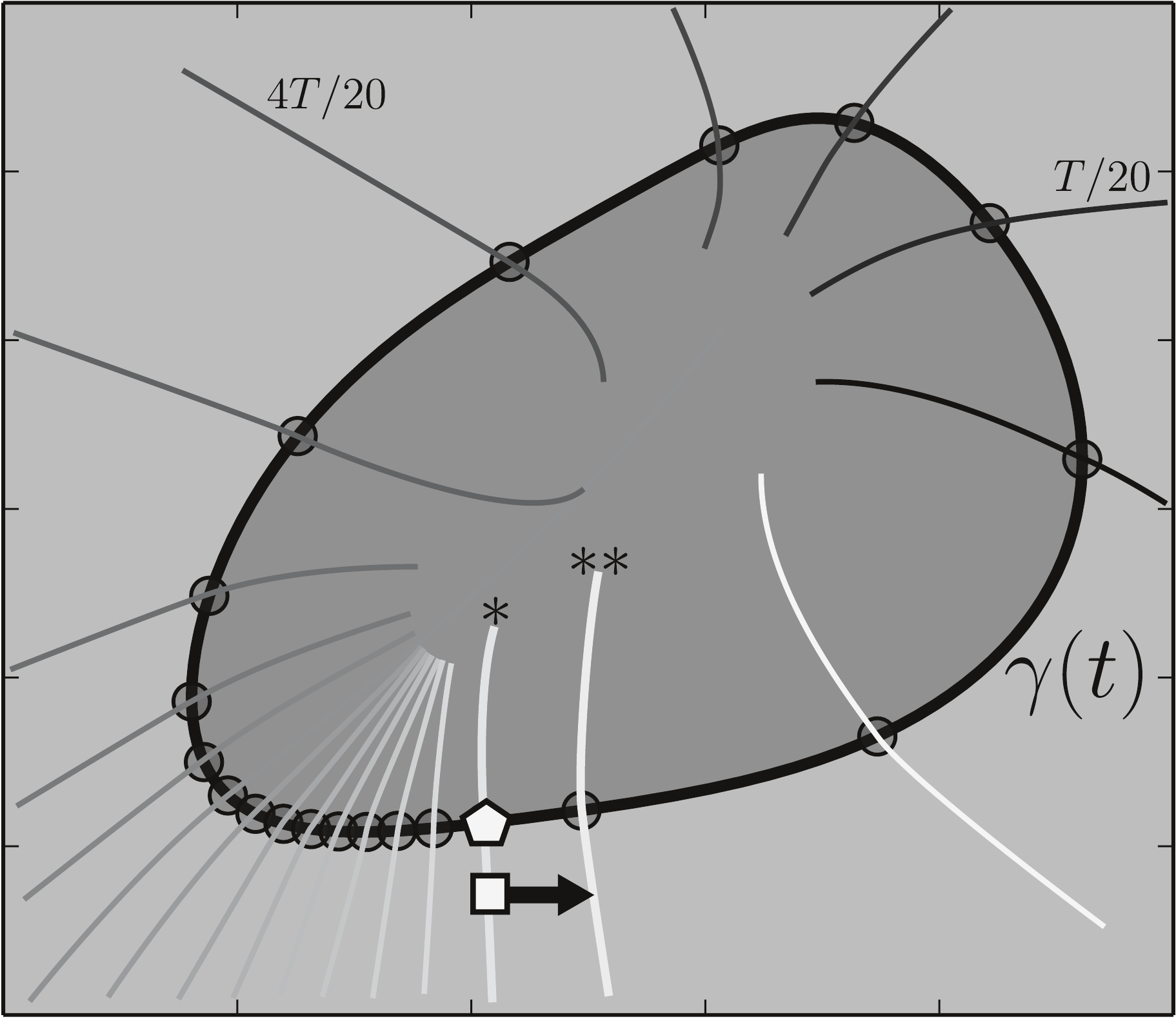}
 \caption{Isochrons in the phase plane of the Morris-Lecar model. The limit cycle (black loop labeled $\gamma(t)$) is marked by circles denoting equally spaced intervals in time, identical to Fig.~\ref{fig:phase_plane}(c). The straight black arrow indicates the effect of an impulse current on the phase of the oscillator, and takes a point on the isochron labeled $*$ (with phase $\theta_{\text{old}}$) to a point on another isochron labeled $**$ (with phase $\theta_{\text{new}}$). The square represents a point in the basin of attraction of the limit cycle $\gamma$, which shares the same asymptotic phase as the point on the limit cycle labeled by a pentagon. Each shaded isochron curve corresponds to the same shade of circle in Fig.~\ref{fig:phase_plane}(c).}
 \label{fig:isochrons}
 \end{figure}

Winfree generalized the notion of phase (which, technically, is only defined on the limit cycle itself) to include all points in the basin of attraction of the limit cycle \cite{winfree_2001}. This generalization begins by choosing an initial condition, say at the square in Fig.~\ref{fig:isochrons}. As time advances in multiples of the limit cycle period $T$, this point converges along the white curve labeled $*$ to a unique point (pentagon) on the limit cycle $\gamma(t)$. The initial condition is then assigned the phase of this unique limit cycle point, $17 T/20$, which we call $\theta_{\text{old}}$. We repeat this method to assign a phase value to every point that converges to the limit cycle.

In mathematical terms, we choose two initial conditions, one in the basin of attraction and another on the limit cycle, $x(0)$ and $y(0)$, respectively. Since $y(0)$ is on the limit cycle, it has some phase associated with it, say $\theta \equiv \theta_{\text{old}}$ (we use the same phase value as above for convenience). If this choice of initial conditions satisfies the property
\begin{equation}\label{eq:asymptotic_phase}
 \lim_{t\rightarrow \infty} \| x(t) - y(t) \| = 0,
\end{equation}
then $x(0)$ is said to have the asymptotic phase $\theta$. The set of all initial conditions sharing this asymptotic phase is called an isochron, and this isochron forms a curve in the plane, labeled $*$ in Fig.~\ref{fig:isochrons}. This idea extends to all other phase values: for each phase value there exists a curve of initial conditions in the basin of attraction satisfying Eq.~\eqref{eq:asymptotic_phase}. Collectively, isochrons form non-overlapping lines in the basin of attraction. The notion of isochrons extends beyond planar limit cycles to limit cycles in any dimension \cite{gucken75}.
 

Equivalently, if $\theta(x)$ denotes the asymptotic phase of the point $x$ in the basin of attraction, then the level curves of $\theta(x)$ are the isochrons. Due to this close relationship between asymptotic phase and isochrons, the terms are used interchangably.

\section{Phase Response Curve}

A fundamental measurement underlying the study of synchrony of coupled oscillators is the phase response curve (PRC): the change in spike timing, or the change in phase, of an oscillating neuron in response to voltage perturbations. If the new phase is denoted $\theta_{\text{new}}$ and the old phase $\theta_{\text{old}}$, then we can quantify the phase shift as
\begin{equation}
 \Delta (\theta_{\text{old}}) = \theta_{\text{new}} - \theta_{\text{old}}.
\end{equation}
This phase shift defines the PRC, and is an easily measurable property of a neural oscillator in both theory and experiment \cite{ermentrout_1996,torben2010}. Neuroscientists often measure the PRC of a neuron by applying a brief current and measuring its change in spike timing. If $\Delta(\theta)$ is negative, then the perturbation lengthens the time to the next spike (phase delay). If $\Delta(\theta)$ is positive, then the perturbation decreases the time to the next spike (phase advance).

In the limit of weak and brief perturbations, the PRC becomes the infinitesimal phase response curve (iPRC). The theory of infinitesimal PRCs was independently proposed by Malkin \cite{malkin1949,malkin1956} and Winfree \cite{winfree_2001}. The iPRC is a result of a Taylor expansion of the phase function,
\begin{equation}
 \theta(\gamma(t) + \varepsilon \eta) = \theta(\gamma(t)) + \varepsilon \nabla \theta(\gamma(t)) \cdot \eta + O(\varepsilon^2),
\end{equation}
where $\eta$ is an arbitrary unit vector direction. The change in phase for this small perturbation is
\begin{equation}
 \Delta \theta = \theta(\gamma(t) + \varepsilon \eta) - \theta(\gamma(t)) = \varepsilon \nabla \theta(\gamma(t)) \cdot \eta + O(\varepsilon^2).
\end{equation}
By taking $\lim_{\varepsilon \rightarrow 0} \Delta \theta /\varepsilon$, we arrive at the expression of the iPRC given a perturbation in the direction $\eta$:
\begin{equation}\label{eq:iprc_derivation}
 \lim_{\varepsilon \rightarrow 0} \frac{\Delta \theta}{\varepsilon} = \nabla \theta(\gamma(t)) \cdot \eta \equiv z(t) \cdot \eta.
\end{equation}
The iPRC is closely related to the PRC. If one finds the PRC by taking small magnitude perturbations and divides the PRC by this small magnitude, then we obtain an approximation to the iPRC. The smaller the magnitude, the better the approximation.

Note that the iPRC $z(t) = (z_1(t), z_2(t), \ldots, z_n(t))$ is a vector in $n$ dimensions, where the $i^{th}$ coordinate represents the iPRC of a perturbation in that coordinate direction. Neuroscientists are often interested in perturbations to the voltage state variable. Assuming voltage lies in the first coordinate, we take $\eta = (1,0,0,\ldots,0)$ and the dot product in Eq.~\eqref{eq:iprc_derivation} recovers the first coordinate $z_1(t)$, which is the iPRC of the voltage variable.

\subsection{Phase Response Curves and Membranes}
\begin{figure}[h!]
\centering
\includegraphics[width=\textwidth]{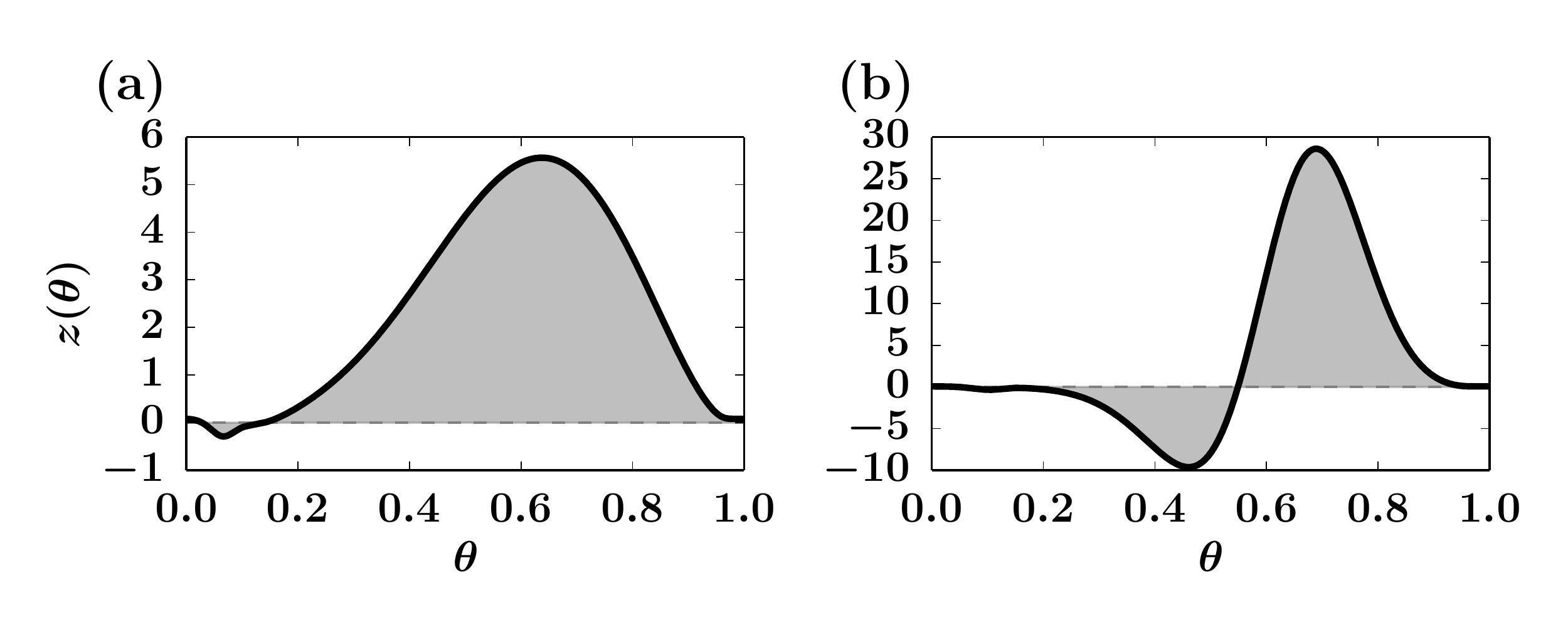}
 \caption{PRC Type I (left), PRC Type II (right) of the Morris-Lecar model. The shaded regions denote the area under the curve. The Type I PRC has a substantial portion of its area on one side of the $x$-axis, while the Type II PRC has substantial portions of its area above and below the $x$-axis.}
 \label{fig:ti_vs_tii_prc}
\end{figure}

The shape of the PRC is informative about the oscillators response to perturbations. An oscillator with a strictly positive (negative) PRC will only ever advance (retard) the phase in response to perturbations, as shown in Fig.~\ref{fig:ti_vs_tii_prc}(a). This type of PRC is classified as a Type I \cite{izhikevich2007}. In neurons, this idea corresponds to advancing or retarding the time to the next spike. On the other hand, a PRC that can simultaneously advance or retard the phase depending on the arrival time of the perturbation is classified at Type II, as shown in Fig.~\ref{fig:ti_vs_tii_prc}(b) \cite{izhikevich2007}. Oscillators with Type II PRC have greater propensity to synchronize to an incoming pulse train because it can both advance and retard its phase.

Membrane oscillations were characterized into two classes by Hodgkin \cite{hodgkin1948,izhikevich2007}: Class I and Class II, as noted in the introduction of this paper.  \cite{jrbe98} showed that Hodgkin's classification could be related to the bifurcation mechanism by which the neurons made the transition from rest to repetitive firing as the input current changed. They showed that Class I excitability corresponds to a SNLC bifircation and Class II to a Hopf bifurcation.  

Remarkably, each PRC type is associated with a distinct excitable membrane property. In \cite{ermentrout_1996,brown_etal_2004}, they show that Class I membranes have Type I PRCs, and Class II membrane oscillations arising from a super- or sub-critical Andronov-Hopf bifurcation have Type II PRCs. 

The figure used to demonstrate Type I and Type II PRCs is derived from the Morris-Lecar model. The parameters used for these models may be found in \cite{ermentrout_terman_2010}.

If the input current is chosen sufficiently far from the onset of oscillations such that membrane oscillations persist, Class I (Class II) oscillators do not generally have Type I (Type II) PRCs. For the reminder of this chapter, we choose parameters close to the onset of Class I (Class II) oscillations. Therefore, any mention of Class I (Class II) oscillations can be assumed to have an associated Type I (Type II) PRC.

\section{Two Weakly Coupled Oscillators}
With the PRC in hand, we now turn to the issue of coupling oscillators into a network.  Networks of neurons that are conductance-based, such as the Morris-Lecar model are generally coupled by synampses and the effects of these synapses is additive, as they are physically currents. Thus, in order to analyze dynamics of networks of rhythmic neurons, we have to (1) derive the interactions that arise after we reduce them to a phase model and (2) see how these interactions depend on the nature of the coupling.  We will study this for a small network of 2, keeping in mind that the pairwise interactions are all that we need in order to simulate and analyze large networks since the networks are formed from weighted sums of the pairwise interactions.  

For pairwise interactions, a natural question to ask is whether or not two oscillators with similar frequencies will synchronize or converge to some other periodic patterns such as ``anti-phase'' where they oscillate a half-cycle apart. It is possible to predict synchrony between coupled oscillators with very general assumptions on the form of coupling by using the phase-reduction technique that we outline below. This generality comes at a price: we must assume that the interactions are ``weak''; that is, the effects of all the inputs to an oscillator are small enough that it stays close to its uncoupled limit cycle.  For this reason, what we next present is often called the theory of weakly-coupled oscillators. 
To make the mathematics easier, we assume the reciprocally coupled oscillators are identical except for the coupling term. That is, we assume coupling of the form,
\begin{equation}\label{eq:odecoupled}
\begin{split}
 \frac{dX_1}{dt} = F(X_1) + \varepsilon G_1(X_2,X_1),\\
 \frac{dX_2}{dt} = F(X_2) + \varepsilon G_2(X_1,X_1),
\end{split}
\end{equation}
where $0<\varepsilon \ll 1$ is small, $X_1,X_2$ are vector valued, and $G_1,G_2$ represent neural coupling. Note that the vector field $F$ is the same in both ODEs, and with $\varepsilon=0$, the stable periodic solution $\gamma(t)$ of $dX/dt = F(X)$ also satisfies both ODEs. To make predictions regarding synchronization, we follow the geometric approach by Kuramoto \cite{kuramoto_1984,ermentrout_terman_2010}.

Let $j=1,2$, $k=3-j$, and $0<\varepsilon\ll 1$. We start with a change of coordinates along the limit cycle, $\theta_j = \Theta(X_j)$, where $\Theta$ is the asymptotic phase function. Because $X_j$ is a function of time, we apply the chain rule to rewrite $\frac{d\theta_j}{dt}$:
\begin{equation*}
 \frac{d\theta_j}{dt} = \nabla_X\Theta(X_j) \cdot \frac{dX_j}{dt}.
\end{equation*}
We substitute $dX_j/dt$ with its vector field definition to yield
\begin{equation*}
 \frac{d\theta_j}{dt} = \nabla_X\Theta(X_j) \cdot F(X_j(t)) + \varepsilon \nabla_X\Theta(X_j) \cdot G_j(X_k,X_j).
\end{equation*}
Finally, we use the normalization property, $\nabla_X\Theta(X_j(t)) \cdot \frac{X_j}{dt} = 1$, where $X_j$ is a periodic solution \cite{ermentrout_terman_2010}.  We arrive at an exact equation that provides some intution of the role of the coupling term $G_j$ and iPRC $\nabla_X\Theta(X_j(t)) \equiv z(t)$ on the phase model $\theta_j$:
\begin{equation}\label{eq:dth_dt_non_instantaneous}
 \frac{d\theta_j}{dt} = 1 + \varepsilon \nabla_X \Theta(X_j) \cdot G_j(X_k,X_j).
\end{equation}
Intuitively, this equation says the phase of the oscillator advances at the usual rate of $d\theta_j/dt = 1$ with an additional weak nonlinear term that depends on the iPRC and the coupling term. We remark that the iPRC term, which is derived by considering instaneous perturbations, appears naturally in a context where perturbations to the phase are not necessarily instaneous.

While Eq.~\eqref{eq:dth_dt_non_instantaneous} is exact, we do not know the form of the solution $X_j$ and therefore can not evaluate this ODE. However, if $\varepsilon$ is sufficiently small, then interactions between the two oscillators are weak and the periodic solutions $X_j(t)$ are almost identical to the unperturbed limit cycle $\gamma(t)$, which is in turn almost identical to $\gamma(\theta_j)$. Making this substitution results in an equation that only depends on the phases $\theta_{1,2}$:
\begin{equation}\label{eq:dth_dt_1}
 \frac{d\theta_j}{dt} = 1 + \varepsilon \nabla_X \Theta[\gamma(\theta_j)]\cdot G_j[\gamma(\theta_k),\gamma(\theta_j)].
\end{equation}
By subtracting off the rotating frame using the change of variables $\phi_j = \theta_j - t$, we can study the effects of coupling without keeping track of a term that grows linearly in time. Eq.~\eqref{eq:dth_dt_1} becomes
\begin{equation*}
 \frac{d\phi_j}{dt} = \varepsilon \nabla \Theta [\gamma(t + \phi_j)] \cdot G_j[\gamma(t + \phi_k),\gamma(t + \phi_j)].
\end{equation*}
All terms that are multiplied by $\varepsilon$ are $T-$periodic so that we can apply the averaging theorem \cite{gh83} to eliminate the explicit time-dependence. (This theorem says that the equation $x'=\varepsilon f(x,t)$ where $f(\cdot,t+T)=f(\cdot,t)$, then, the dynamics of $x$ are close to those of $y$ for $\varepsilon$ small, where $y'=\bar{f}(y)$ and $\bar{f}(y)=(1/T)\int_0^Tf(y,t)\ dt.$) We average the right-hand sides over one cycle to obtain:
\begin{equation}
\label{eq-phase}
\begin{split}
\frac{d\phi_j}{dt} &= \varepsilon H_j(\phi_k-\phi_j),\\
H_j(\psi) &= \frac{1}{T}\int_0^T \nabla \Theta[\gamma(t)]\cdot G_j[\gamma(t+\psi),\gamma(t)]\ dt.
\end{split}
\end{equation}
That is, we have reduced the system of two-coupled $n-$dimensional systems to a pair of coupled scalar equations. It should be clear that if the coupling terms are additive (as they would be in the case of synaptic coupling) and there are $m$ coupled oscillators, then the phase equations will have the general form:
\begin{equation}
\label{eq:phase-gen}
\frac{d\phi_j}{dt} = \varepsilon \sum_{k=1}^m H_{jk}(\phi_k-\phi_j), \quad j=1,\ldots,m.
\end{equation}
We can make one more reduction in dimension by observing that all the interactions in (\ref{eq:phase-gen}) depend only on the phase-difference. Thus we can study the {\em relative} phases by setting $\psi_j=\phi_j-\phi_1$, for $j=2,\ldots,m$ and obtain the $m-1-$dimensional set of equations:
\begin{equation}
\label{eq:relphs}
\frac{d\psi_j}{dt} = \varepsilon \left(\sum_{k=1}^m H_{jk}(\psi_k-\psi_j)-H_{1k}(\psi_k)\right),\quad j=2,\ldots,m.
\end{equation}
where we set $\psi_1=0$. The beauty of these equations is that equilibrium points correspond to periodic solutions to the original set of coupled oscillators and these periodic solutions have the same stability properties as the equilibria of (\ref{eq:relphs}).  For example, synchrony of the coupled oscillators would correspond to a solution to (\ref{eq:relphs}) where $\psi_2=\ldots=\psi_m=0.$   An easily computed sufficient condition for stability of equilibria of (\ref{eq:relphs}) can be found in \cite{be92}. For the remainder of this chapter, we focus on $m=2$, and define $\psi=\phi_2-\phi_1$ to obtain a single scalar equation for the phase-difference of the two oscillators: 
\begin{equation}\label{eq:phase_model}
 \frac{d\psi}{dt}=\varepsilon [H_2(-\psi)-H_1(\psi)] \equiv \varepsilon H(\psi),
\end{equation}
where
\begin{equation}\label{eq:hfun}
 H_j(\psi) = \frac{1}{T}\int_0^T \! z(t) \cdot G_j(\gamma(t+\psi), \gamma(t))\, \mathrm{d}t
\end{equation}
and $z(t)=\nabla\Theta(\gamma(t))$ as above. 
The function $H_j $ is often called the interaction function \cite{schwemmer_lewis2012} and is the convolution of the coupling term $G_j$ with the iPRC $z$. 

\noindent {\bf Remark 1.} We note that equation (\ref{eq:phase_model}) was derived under the assumption that there were no frequency difference the two oscillators.  However, if the frequency difference are small, that is, $O(\varepsilon)$, then the equations (\ref{eq:phase-gen}) have an addional constant term, $\varepsilon \omega_j$ representing the uncoupled frequency difference from that of $\gamma(t).$  In neural models, the easiest way to change the frequency is by adding some additional current, $\delta I$. In this case
\begin{equation}
\label{eq:freq}
\omega = (\delta I /C_m)  \frac{1}{T} \int_0^T z_V(t)\ dt,
\end{equation}
where $z_V(t)$ is the voltage component of the iPRC and $C_m$ is the membrane capacitance.  Equation (\ref{eq:freq}) is intuitively appealing: oscillators with positive PRCs are the most sensitive to currents since their average will generally be larger than PRCs that have both positive and negative values. When the oscillators have slightly different frequencies (in the sense of this remark), then equation (\ref{eq:phase_model}) becomes:
\begin{equation}
\label{eq:fd}
\frac{d\psi}{dt} = \varepsilon [ \mu + H(\psi)]
\end{equation}  
where $\mu=\omega_2-\omega_1.$ Frequency differences have the effect of shifting the interaction function up and down.

\begin{figure}[h!]
\centering
 \includegraphics[width=.95\textwidth]{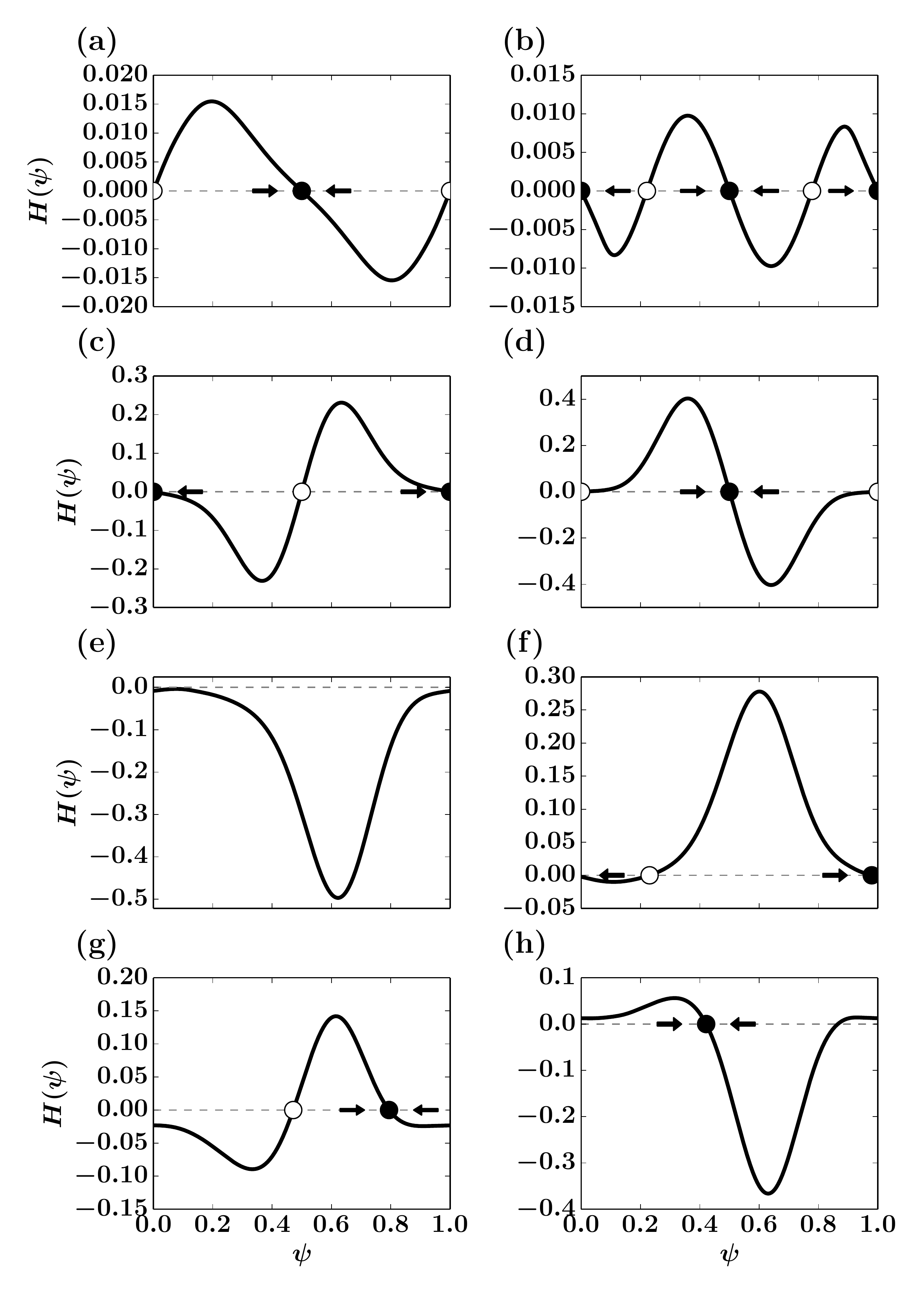}
 \caption{Stability analysis of the right-hand-side function $H$ of Eq.~\eqref{eq:phase_model}. (a) Class I excitatory to Class I excitatory coupling. (b) Class I inhibitory to Class I inhibitory coupling. (c) Class II excitatory to Class II excitatory coupling. (d) Class II excitatory to Class II excitatory coupling. (e) Class I inhibitory to Class II excitatory coupling. (f) Class I excitatory to Class II inhibitory coupling. (g) Class I excitatory to Class II excitatory coupling. (h) Class I inhibitory to Class II inhibitory coupling. The $x$-axis of each figure is marked by fractions of the corresponding period.}
 \label{fig:hfun_combined}
\end{figure}

\noindent {\bf Remark 2.} Before continuing with our discussion of the behavior of the phase models, we want to briefly discuss the issues that arise from coupling different oscillataors together (e.g. a class I with a class II, such as figure \ref{fig:hfun_combined}e,f).   Our results for phase models are strictly valid when the uncoupled systems are identical. However, in coupling Class I,II neurons, the uncoupled oscillators are different and so, the limit cycles are not the same functions.  Thus the equations presented for the interaction functions (\ref{eq:hfun}) are not correct.  We can still apply the averaging theorem as long as we adjust parameters of the two distinct systems so that the uncoupled frequencies are identical. We can then use the same ideas to compute the interaction functions. Let $\gamma_{j,k}(t),z_{j,k}(t)$ be the limit cycles and iPRCs of the two uncoupled systems. By assumption, they are both $T-$periodic.  Then:
\begin{eqnarray*}
H_1(\psi)&=& \frac{1}{T}\int_0^T z_1(t) \cdot G_1(\gamma_2(t+\psi),\gamma_1(t))\ dt \\
H_2(\psi)&=& \frac{1}{T}\int_0^T z_2(t) \cdot G_2(\gamma_1(t+\psi),\gamma_2(t))\ dt
\end{eqnarray*}
With these changes for the heterogenous oscillators, we can now proceed.

There are many advantages to the result in Eq.~\eqref{eq:phase_model}. The ODE is autonomous and scalar, so we can apply a standard stability analysis on the phase line. Fixed points on the phase line correspond precisely to stable (unstable) phase locked solutions. In particular, a  fixed point at $\psi = 0$ corresponds to synchrony, and a fixed point at $\psi=T/2$ corresponds to anti-synchrony. 

We show various examples of the right-hand-side function $H$ in Fig.~\ref{fig:hfun_combined} using synaptically coupled Morris-Lecar models (relevant parameters are listed in Table \ref{tbl:ml_parms_syn}). The phase line is shown on the $x$-axis of each subfigure. On the phase line, a black filled circle (open circle) corresponds to an asymptotically stable (unstable) phase locked solution. Fig.~\ref{fig:hfun_combined}(a) is of two weakly coupled Class I neurons with reciprocal excitatory coupling. In this case, the phase model predicts that coupled oscillators will asymptotically converge to an anti-phase rhythm. If there is more than one stable solution, as in Fig.~\ref{fig:hfun_combined}(b), then the asymptotic phase difference depends on the initial relative phase shift of the oscillators. Initializing with a sufficiently small phase difference results in asymptotic synchrony, while initializing with a larger phase difference close to half a period results in asymptotic anti-synchrony. This subfigure corresponds to two weakly coupled Class I neurons with reciprocal inhibitory coupling.

The remainder of Fig.~\ref{fig:hfun_combined} considers Class II excitatory to Class II excitatory coupling (Fig.~\ref{fig:hfun_combined}(c)), Class II inhibitory to Class II inhibitory coupling (Fig.~\ref{fig:hfun_combined}(d)), Class I inhibitory to Class II excitatory coupling (Fig.~\ref{fig:hfun_combined}(e)), Class I excitatory to Class II inhibitory coupling (Fig.~\ref{fig:hfun_combined}(f)), Class I excitatory to Class II excitatory coupling (Fig.~\ref{fig:hfun_combined}(g)), and Class I inhibitory to Class II inhibitory coupling (Fig.~\ref{fig:hfun_combined}(e)). As in the preceding discussion, determining asymptotic stability is a straightforward matter of finding stable fixed points on the phase line.

In Fig.~\ref{fig:hfun_combined}(e), there are no fixed points. Such a case corresponds to phase drift: the oscillators never phase-lock. Reciprocal coupling of Class I with Class II neurons is tricky because we must choose the frequencies to be sufficiently similar (see Remark 2, above). We find that choosing $I=43.5$ for the Class I neuron and $I=88.5$ for the Class II neuron both preserves the salient features of the respective PRC types and results in good agreement in oscillator frequency. Why are there no fixed points in case (e) and why are the fixed points nearly degenerate in (f)? We can understand this as follows. From Remark 2, we see that $H_1$ is found by convolving a Type I PRC with an excitatory synapse.  This will result in $H_1$ positive everywhere.  On the other hand, $H_2$ is found by convolving a Type II PRC with an inhibitory synapse that results in a mixture of positive and negative regions. The large positive $H_1$, when subtracted from $H_2$ to get the equation for the phase-difference will be negative as seen in the figure. More intuitively, the excitatory synapse onto the the Type I neuron will constantly {\em advance} the phase of neuron 1 while the inhibitory synapse will cause a mix of advance and delay as the PRC is Type II. This there will be a net advance of oscillator 1 over oscillator 2 and we will see drift ($H(\psi)<0$ everywhere).  Similar considerations hold for panel (f).   
From Remark 1 (above), we recall that by introducing small frequency differences, we can shift the interaction functions up and down. Thus, we could get a phase-locked solution, in, e.g., panel (f) by adding a small depolarizing curren to oscillator 2, thus allowing it to speed up.

\begin{figure}[h!]
\includegraphics[width=\textwidth]{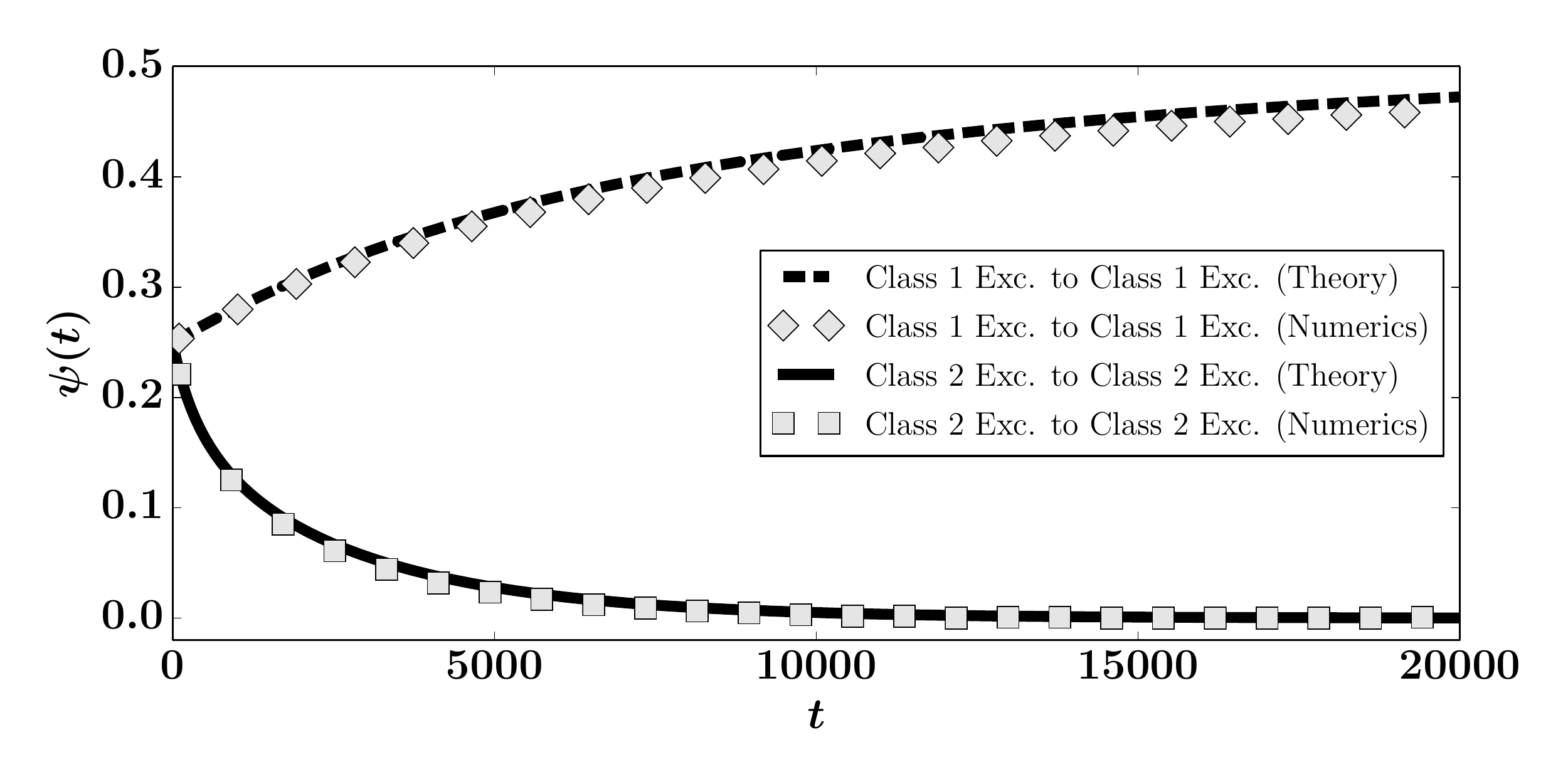}
 \caption{Difference in synchronization rates between Class I excitatory (dashed black) and Class II excitatory (solid black) reciprocally coupled Morris-Lecar oscillators. The diamonds and squares represent numerical phase differences for Class I and Class II reciprocal coupling, respectively. The $y$-axis is labeled by the fraction of the respective periods.}\label{fig:theory_vs_numerics}
\end{figure}

We list additional observations that follow from Eq.~\eqref{eq:phase_model}.
\begin{itemize}
 \item If the interaction terms $G_i$ are delta functions (used for arbitrarily fast synapses), the interaction function $H_j$ is directly proportional to the PRC.
 \item If reciprocal coupling is the same and the uncoupled  oscillators are the same, then $G_1 = G_2$, and $H_1 = H_2 \equiv H$ and the right hand side of Eq.~\eqref{eq:phase_model} is proportional to the odd part of $H$, denoted $H_{\text{odd}}$:
\begin{equation}\label{eq:neg2hodd}
 \frac{d\psi}{dt}=-\varepsilon 2 H_{\text{odd}}(\psi).
\end{equation}
 \item In addition to predicting asymptotic stability, Eq.~\eqref{eq:phase_model} also provides convergence rates of solutions, and therefore synchronization rates of the full dynamics.
\end{itemize}
We demonstrate the accuracy of the convergence rates in Fig.~\ref{fig:theory_vs_numerics}. The dashed and solid curves are computed using Eq.~\eqref{eq:hfun} with parameters chosen to represent Class I excitatory to Class I excitatory coupling and Class II excitatory to Class II excitatory coupling, respectively. The diamonds and squares represent the numerical phase difference in the full model. We find the phase difference of the full model by computing spike timing differences following the numerical integration of Eqs.\eqref{eq:ml_rhs}--\eqref{eq:ml_syn}. We choose $\varepsilon=0.0025$, which is sufficiently small for accurate predictions.

\section{Summary of Reciprocal Coupling}
The results are summarized in Table \ref{tbl:coupling}, where the headers denote the excitatory or inhibitory effect of a given neuron. ``Class I excitatory'' is shorthand for an excitatory Class I neuron, and ``Class II excitatory'' is shorthand for an excitatory Class II neuron. Synaptic driving potential is $0,-75$mV, for excitatory and inhibitory synaptic coupling, respectively.

\begin{table}[h!]
\caption{Survey of asymptotically stable convergence. Each number represents the phase-locked solution as a fraction of the total period. A table entry with two numbers implies the existence of two phase-locked solutions. Horizontal dashes denote phase drift.}\label{tbl:coupling}
\centering
\begin{tabular}{|cc|c|c|c|c|}
\cline{3-6}
\hline & & \multicolumn{2}{ c| }{Class I} & \multicolumn{2}{ c| }{Class II} \\ \cline{3-6}
& & Excitatory & Inhibitory & Excitatory & Inhibitory \\ \cline{1-6}
\multicolumn{1}{ |c  }{\multirow{2}{*}{CI} } &
\multicolumn{1}{ |c| }{Ex} &  0.5 & - & 0.210 & 0  \\ \cline{2-6}
\multicolumn{1}{ |c  }{}                        &
\multicolumn{1}{ |c| }{In} & \cellcolor{black!60} &0.5,\,\,0 &  - &   0.5 \\ \cline{1-6}
\multicolumn{1}{ |c  }{\multirow{2}{*}{CII} } &
\multicolumn{1}{ |c| }{Ex} & \cellcolor{black!60} & \cellcolor{black!60} &   0 & 0.862   \\ \cline{2-6}
\multicolumn{1}{ |c  }{}                        &
\multicolumn{1}{ |c| }{In} & \cellcolor{black!60} & \cellcolor{black!60} & \cellcolor{black!60} &   0.5 \\ \cline{1-6}
\end{tabular}
\end{table}
Numbers in the table denote phase locked solutions as a proportion of the respective period. As mentioned earlier, parameter values for Class I, Class II neurons and excitatory, inhibitory synapses are chosen according to Table \ref{tbl:ml_parms_syn}. Asymptotic convergence to $0$ corresponds to synchrony, while convergence to $0.5$ corresponds to anti-phase.

The diagonal entries of the table as well as the four combinations of Class I to Class II excitatory/inhibitory coupling have been shown in Fig.~\ref{fig:hfun_combined}. The remaining table entries consider Class I excitatory to Class I inhibitory coupling (phase drift), and Class II excitatory to Class II inhibitory coupling (phase locked at 0.862).

\section{Conclusion}

Reducing tonically firing neurons to a phase model allows us to formulate a mathematically precise phase description of neural synchronization. Using this phase description, we quantified perturbations of phase using phase response curves. We also demonstrated using a qualitative geometric argument how a perturbation can push solutions to different isochrons, resulting in a phase shift. The phase description of a neural oscillator is useful because just one scalar variable represents the dynamics of what are generally high dimensional systems involving many conductances.

The knowledge of the iPRC and the coupling term is useful in predicting the synchronization outcome. By convolving the iPRC with the coupling term(s), we derive an autonomous scalar differential equation for the phase difference dynamics, which faithfully reproduces synchronization in the full numerical integration. Moreover, because the phase difference dynamics is given by a scalar, autonomous differential equation, an analysis on the phase line provides all the necessary insight into asymptotic phase-locking. We use a phase-line analysis to predict synchronization of various reciprocally coupled oscillators. Our synapses are slow, but the observations happen to agree with what is known in the literature for fast synapses, in particular that Class I excitatory to Class I excitatory coupling tends not to synchronize at zero lag, while Class II excitatory to Class II excitatory coupling tends to synchronize \cite{hansel_etal_1995}.

In addition to predicting asymptotic phase-locked states, knowledge of the iPRC and coupling term also leads to predictions of synchronization rates, as shown in Fig.~\ref{fig:ti_vs_tii_prc}. This figure also demonstrates the flexibility of weak coupling theory. Despite the nonlinear nature of synaptic coupling, sufficiently weak interactions leads to accurate predictions of both rates and stability of phase-locking. 

Weak coupling theory naturally applies to networks of $N$ coupled oscillators with virtually no modifications \cite{ermentrout_terman_2010}, and relevant applications arise in biology, chemistry, and physics. Examples include swimming locomotion of dogfish, lamprey and lobster \cite{ermentrout_kopell_1986}, communication of fireflies \cite{hoppensteadt_izhikevich_2012}, reaction-diffusion chemical reactions \cite{kuramoto_1984}, coupled reactor systems \cite{neu_coupled_1979}, and Josephson junctions \cite{watanabe_strogatz_1994}.

\appendix

\section{Morris-Lecar Model}
The Morris-Lecar model is a planar conductance-based model, originally developed to model various oscillations observed in barnacle muscle \cite{morris_lecar_1981}. Using notation in Eq.~\eqref{eq:odemain}, we let $X=[V,w]^T$
\begin{equation}\label{eq:ml_rhs}
F(X)= 
 \begin{bmatrix}
  (I - g_L(V-V_L) - g_{Ca}m_{\infty}(V)(V-V_{Ca}) - g_k w (V-V_k))/C\\
  (w_{\infty}(V)-w)/\tau_w(V)
 \end{bmatrix},
\end{equation}
where
\begin{equation}\label{eq:ml_rhs2}
\begin{split}
m_{\infty}(V) &= \left(1+\tanh\left[\frac{V-V_1}{V_2}\right]\right)/2,\\
w_{\infty}(V) &= \left(1+\tanh\left[\frac{V-V_3}{V_4}\right]\right)/2,\\
\tau_w(V)&=1/\left(\phi \cosh\left[\frac{V-V_3}{2V_4}\right]\right),
\end{split}
\end{equation}
are defined to keep the notation compact. The model is not analytically tractable due to multiple nonlinearities, so we proceed numerically.

\subsection{Synaptic Dynamics}\label{A:ml_syn}

We use the following coupling functions in our numerical examples of Eq.~\eqref{eq:phase_model}. Let $X_i = [V_i,w_i,s_i]^T$. The coupling terms are defined as
\begin{equation}\label{eq:ml_g}
G_1(X_2,X_1)= 
 \begin{bmatrix}
  g s_2 (V_{syn}-V_1)\\
  0\\
  0
 \end{bmatrix}, \quad
 G_2(X_1,X_2)= 
 \begin{bmatrix}
  g s_1 (V_{syn}-V_2)\\
  0\\
  0
 \end{bmatrix},
\end{equation}
where the dynamics of $s_i$ satisfy
\begin{equation}\label{eq:ml_syn}
\begin{split}
s_i'&=\alpha k(V) (1-s_i)-\beta s_i, \quad i=1,2,\\
k(V)&=\frac{1}{1+\exp(-(V-v_t)/v_s)}.
\end{split}
\end{equation}

\begin{table}[h!]
\numberwithin{table}{section}
\caption{Synaptic Coupling Parameter Values}\label{tbl:ml_parms_syn}
\centering
\begin{tabular}
{ l l }
 Parameter 	  & Value(s)\\
 \hline
  $\alpha$ 	  & 1\\
  $\beta$ 	  & 0.05 \\
  $V_t$ 	  & -1.2 \\
  $V_s$ 	  & 2 \\
  $V_{syn}$	  & $0$mV,$-75$mV\\
  $g$		  & 5\\
\end{tabular}
\end{table}
These dynamics are often used to model synaptic interactions. Qualitatively, the rate of activation $s_i$ is determined by $\alpha$ and the voltage-dependent degree of activation $k(V)$. If voltage is large, say from an action potential, and the synapse is inactive, $k$ and $(1-s_i)$ are maximized, resulting in an increase in synaptic activity. Eventually, the synapse is maximally active, and the voltage has returned to its resting state, so $k(V)$ is minimized close to zero and the synaptic activity decays at a rate $\beta$.

We choose $V_{syn} = 0$mV, $-75$mV for excitatory and inhibitory coupling, respectively.

\bibliographystyle{plain}
\bibliography{./biblio}

\begin{thebibliography}{10}

\bibitem{brown_etal_2004}
Moehlis~J Brown, E and P~Holmes.
\newblock On the {Phase} {Reduction} and {Response} {Dynamics} of {Neural}
  {Oscillator} {Populations}.
\newblock {\em Neural Computation}, 16(4):673--715, 2004.

\bibitem{ermentrout_1996}
Bard Ermentrout.
\newblock Type {I} {Membranes}, {Phase} {Resetting} {Curves}, and {Synchrony}.
\newblock {\em Neural Computation}, 8(5):979--1001, 1996.

\bibitem{be92}
G~Bard Ermentrout.
\newblock Stable periodic solutions to discrete and continuum arrays of weakly
  coupled nonlinear oscillators.
\newblock {\em SIAM Journal on Applied Mathematics}, 52(6):1665--1687, 1992.

\bibitem{ermentrout_terman_2010}
GB~Ermentrout and DH~Terman.
\newblock {\em Mathematical {Foundations} of {Neuroscience}}, volume~35 of {\em
  Interdisciplinary {Applied} {Mathematics}}.
\newblock Springer New York, New York, NY, 2010.

\bibitem{fell_axmacher_2011}
J~Fell and N~Axmacher.
\newblock The role of phase synchronization in memory processes.
\newblock {\em Nature Reviews Neuroscience}, 12(2):105--118, 2011.

\bibitem{fries_2005}
P~Fries.
\newblock A mechanism for cognitive dynamics: neuronal communication through
  neuronal coherence.
\newblock {\em Trends in Cognitive Sciences}, 9(10):474--480, 2005.

\bibitem{gucken75}
John Guckenheimer.
\newblock Isochrons and phaseless sets.
\newblock {\em Journal of Mathematical Biology}, 1(3):259--273, 1975.

\bibitem{gh83}
John Guckenheimer and Philip Holmes.
\newblock {\em Nonlinear oscillations, dynamical systems, and bifurcations of
  vector fields}, volume~42.
\newblock Springer Science \& Business Media, 1983.

\bibitem{hansel_etal_1995}
D~Hansel, G~Mato, and C~Meunier.
\newblock Synchrony in excitatory neural networks.
\newblock {\em Neural computation}, 7(2):307--337, 1995.

\bibitem{hodgkin1948}
AL~Hodgkin.
\newblock The local electric changes associated with repetitive action in a
  non-medullated axon.
\newblock {\em The Journal of physiology}, 107(2):165--181, 1948.

\bibitem{hoppensteadt_izhikevich_2012}
Frank~C Hoppensteadt and Eugene~M Izhikevich.
\newblock {\em Weakly connected neural networks}, volume 126.
\newblock Springer Science \& Business Media, 2012.

\bibitem{izhikevich2007}
EM~Izhikevich.
\newblock {\em Dynamical systems in neuroscience: The Geometry of Excitability
  and Bursting}.
\newblock MIT press, 2007.

\bibitem{ermentrout_kopell_1986}
N~Kopell and GB~Ermentrout.
\newblock Symmetry and phaselocking in chains of weakly coupled oscillators.
\newblock {\em Communications on Pure and Applied Mathematics}, 39(5):623--660,
  1986.

\bibitem{kuramoto_1984}
Yoshiki Kuramoto.
\newblock {\em Chemical oscillations, waves, and turbulence}, volume~19.
\newblock Springer Science \& Business Media, 2012.

\bibitem{malkin1949}
IG~Malkin.
\newblock Methods of {Poincare} and {Liapunov} in theory of non-linear
  oscillations.
\newblock {\em Gostexizdat}, 1949.

\bibitem{malkin1956}
IG~Malkin.
\newblock Some problems in nonlinear oscillation theory.
\newblock {\em Gostexizdat, Moscow}, 541, 1956.

\bibitem{morris_lecar_1981}
C~Morris and H~Lecar.
\newblock Voltage oscillations in the barnacle giant muscle fiber.
\newblock {\em Biophysical journal}, 35(1):193, 1981.

\bibitem{neu_coupled_1979}
J.~Neu.
\newblock Coupled {Chemical} {Oscillators}.
\newblock {\em SIAM Journal on Applied Mathematics}, 37(2):307--315, 1979.

\bibitem{nunez_srinivasan_2006}
PL~Nunez and R~Srinivasan.
\newblock {\em Electric {Fields} of the {Brain}: {The} neurophysics of {EEG}}.
\newblock OUP USA, 2nd edition, 2006.

\bibitem{jrbe98}
John Rinzel and G~Bart Ermentrout.
\newblock Analysis of neural excitability and oscillations.
\newblock {\em Methods in neuronal modeling}, 2:251--292, 1998.

\bibitem{schnitzler_gross_2005}
A~Schnitzler and J~Gross.
\newblock Normal and pathological oscillatory communication in the brain.
\newblock {\em Nature Reviews Neuroscience}, 6(4):285--296, 2005.

\bibitem{schwemmer_lewis2012}
MA~Schwemmer and TJ~Lewis.
\newblock {\em The {Theory} of {Weakly} {Coupled} {Oscillators}}.
\newblock 2012.

\bibitem{torben2010}
Benjamin Torben-Nielsen, Marylka Uusisaari, and Klaus~M Stiefel.
\newblock A comparison of methods to determine neuronal phase-response curves.
\newblock {\em Frontiers in neuroinformatics}, 4, 2010.

\bibitem{watanabe_strogatz_1994}
S~Watanabe and SH~Strogatz.
\newblock Constants of motion for superconducting josephson arrays.
\newblock {\em Physica D: Nonlinear Phenomena}, 74(3):197--253, 1994.

\bibitem{winfree_2001}
AT~Winfree.
\newblock {\em The Geometry of Biological Time}.
\newblock Interdisciplinary Applied Mathematics. Springer New York, 2001.

\end{thebibliography}

\end{document}